\documentstyle[multicol,aps,epsf]{revtex}
\begin{document}

\title{Shear Viscosity of Hot Scalar Field Theory in the Real-Time Formalism}

\author{Enke Wang$^a$ and Ulrich Heinz$^b$}

\address{$^a$Institute of Particle Physics, Huazhong Normal University,
         Wuhan 430079, China\\
         $^b$Department of Physics, The Ohio State University, 174 West 18th
         Avenue, Columbus, OH 43210, USA}

\date{\today}

\maketitle

\begin{abstract}
Within the closed time path formalism a general nonperturbative 
expression is derived which resums through the Bethe-Salpter equation
all leading order contributions to the shear viscosity in hot scalar 
field theory. Using a previously derived generalized 
fluctuation-dissipation theorem for nonlinear response functions in 
the real-time formalism, it is shown that the Bethe-Salpeter equation 
decouples in the so-called $(r,a)$ basis. The general result is applied 
to scalar field theory with pure $\lambda\phi^4$ and mixed 
$g\phi^3{+}\lambda\phi^4$ interactions. In both cases our calculation 
confirms the leading order expression for the shear viscosity previously 
obtained in the imaginary time formalism.
\end{abstract}


\begin{multicols}{2}

\section{Introduction}
\label{sec1}

The investigation of the transport properties of hot, weakly coupled 
relativistic plasmas is helpful for an understanding of the collective 
effects in a Quark Gluon Plasma (QGP) produced in relativistic heavy 
ion collisions. The Kubo formulae of linear response theory \cite{Zubarev} 
provide a way to calculate transport coefficients using Feynman diagram
techniques in thermal field theory. The shear viscosity can be expressed
as\cite{Zubarev,Hosoya}
  \begin{equation}
  \label{eta}
  \eta = -{1\over 10}\int d^3 x'\int_{-\infty}^0 \!\!\!\! dt
       \int_{-\infty}^{t} \!\!\!\! dt'
       \left\langle\pi ^{\mu\nu}(0)
       \pi _{\mu\nu}({\bf x}',t')\right\rangle^{\rm ret}\, ,
  \end{equation}
where $\pi _{\mu\nu}\ (\mu,\nu{=}0,1,2,3)$ is the traceless
viscous-pressure tensor which is imbedded in the energy-momentum
tensor and in the comoving frame has only spatial components,
$\pi_{ij}$. For a scalar field $\pi_{ij} = (\delta_{ik}\delta_{jl}
 -{1\over 3}\delta_{ij}\delta_{kl})\partial_k \phi\,\partial_l\phi,
 (i,j,k,l=1,2,3)$. $\langle\dots\rangle$ stands for the thermal
expectation value, and the superscript ``ret'' denotes the retarded 
2-point function which is related to the spectral function of the 
$\pi\pi$ correlator. Previous studies have shown that even at weak 
coupling a one-loop calculation \cite{Hosoya,Jeon1,Wang1} of the
spectral function of the composite field $\pi_{ij}$ is incomplete:
the same finite thermal lifetime of the particles which is already
required to regulate a pinch singularity in the one-loop result
serves as a regulator for increasingly severe pinch singularities
at higher order. Since the thermal width compensates for explicit 
coupling constant factors arising from the interaction vertices, an 
infinite number of multiloop diagrams is found to contribute at 
leading order to the transport coefficients \cite{Jeon2,Wang2,CHK}. 
A nonperturbative calculation is required to resum this infinite set 
of leading order diagrams.

In two pioneering papers, Jeon \cite{Jeon1,Jeon2} derived within
the Imaginary Time Formalism (ITF) a set of diagrammatic "cutting" 
rules, identified all diagrams which contribute to the viscosity at
leading order, and then summed the geometric series of cut ladder 
diagrams to obtain the viscosity. This calculation was a remarkable
{\em tour de force} which, to our knowledge, has never been repeated
and checked. In this paper we will do so by reformulating the problem
in real time using the Closed Time Path (CTP) formalism 
\cite{Schwinger,Bakshi,Keldysh,Chou}. The present paper provides 
technical details for the short report given in \cite{Wang2} and
generalizes that and other work \cite{CHK} to scalar fields with 
arbitrary interaction Lagrangians. In addition to giving an independent 
recalculation of the leading order shear viscosity in hot scalar field 
theory with 3- and 4-point interactions, the present work also provides
a natural starting point for a future generalization to dynamical 
problems where the plasma is (slightly) out of thermal equilibrium.

In the real-time formalism the resummation of an infinite number of
leading-order contributions to the viscosity is done relatively easily 
by writing down and solving a Bethe-Salpeter (BS) integral equation 
for the 4-point function. With the help of the generalized 
Fluctuation-Dissipation Theorem (FDT) for nonlinear response functions 
\cite{Wang3} and further simplifications arising from the so-called 
``pinch limit'' (see discussion below), the BS equations for different 
thermal components of the 4-point Green function can be decoupled at 
leading order in the coupling constant \cite{Wang2,CHK,CHS}. We use this 
here to derive a general expression for the nonperturbative calculation 
of the leading order contribution to the shear viscosity in scalar field 
theories with arbitrary interactions. Its application to scalar field 
theories with pure $\lambda\phi^4$ and mixed $g\phi^3{+}\lambda\phi^4$ 
interactions is then straightforward and reproduces Jeon's ITF results 
in an economic way. A similar structure of the calculation is expected
for gauge theories in the leading logarithmic approximation 
\cite{AMY}.

This paper is organized as follows: In Sec.~\ref{sec2} we present the 
BS equation which resums all leading order diagrams to the shear
viscosity, and we derive a general expression for its nonperturbative 
calculation. In Sec.~\ref{sec3} we evaluate the kernel of the BS 
integral equation for pure $\lambda\phi^4$ and mixed $g\phi^3{+}\lambda\phi^4$ 
scalar field theories. A short summary is given in Sec.~\ref{sec4}.

\newpage

\section{Resummation of ladder diagrams\\
         and a Nonperturbative expression\\
         for the shear viscosity}
\label{sec2}
\subsection{CTP formalism in the $(r,a)$ basis}
\label{sec2a}

The Kubo formula (\ref{eta}) for the shear viscosity can be expressed 
in terms of the Fourier-transformed traceless stress tensor Wightman 
function \cite{Jeon1,Jeon2},
 \begin{equation}
 \label{eta1}
   \eta = {\beta \over 20}\lim_{p^0,\bbox{p} \to 0}\int d^4x\,
   e^{ip{\cdot}x} \langle\pi_{ij}(x)\pi^{ij}(0)\rangle\, ,
 \end{equation}
where $\beta$ is the inverse temperature. The Wightman function 
$\langle\pi_{ij}(x)\pi^{ij}(0)\rangle$ is identical with the
(12)-component of the matrix propagator for the composite $\pi$ 
field in the CTP formalism. For any field $\psi$ (composite or elementary) 
the four components of this matrix propagator are defined as
 \begin{equation}
 \label{Ga1a2}
   \Delta^{(\psi\psi)}_{a_1 a_2}(x_1,x_2) \equiv -i
   \langle T_p[\psi_{a_1}(x_1)\psi_{a_2}(x_2)]\rangle \, ,
 \end{equation}
where $T_p$ represents the time ordering operator along the closed
time path (corresponding, respectively, to normal and antichronological 
time ordering of operators with time arguments on its upper and lower
branch), and $a_1,a_2{\,\in\,}\{1,2\}$ indicate on which of the two 
branches the $\psi$ fields are located. The scalar field propagator 
will be simply denoted by $\Delta$:
 \begin{equation}
 \label{delta}
   \Delta_{a_1 a_2}(x_1,x_2) \equiv -i
   \langle T_p[\phi_{a_1}(x_1)\phi_{a_2}(x_2)]\rangle \, .
 \end{equation}
Following \cite{Chou} we define
 \begin{equation}
 \label{ar}
  \phi_a(x) = \phi_1(x){-}\phi_2(x) ,\quad
  \phi_r(x) = {\textstyle{1\over 2}} (\phi_1(x){+}\phi_2(x)) 
 \end{equation}
and the 2-point Green function in the $(r,a)$ basis
 \begin{equation}
 \label{Gaf1af2}
  \Delta_{\alpha_1\alpha_2}(x_1,x_2) \equiv -i 2^{n_r-1} 
  \langle T_p[\phi_{\alpha_1}(x_1)\phi_{\alpha_2}(x_2)]\rangle .
 \end{equation}
Here $\alpha_1,\alpha_2{\,\in\,}\{a,r\}$, and $n_r$ is the number of $r$
indices among $\{\alpha_1,\alpha_2\}$. Substituting Eq.~(\ref{ar}) into
(\ref{Gaf1af2}) it is easy to show the following relations for the 
$\phi$ propa\-gator:
 \begin{mathletters}
 \label{d}
 \begin{eqnarray}
 \Delta_{rr}(x_1, x_2) &=&\Delta_{12}(x_1, x_2)+\Delta_{21}(x_1,x_2)
   \nonumber\\
   &=&-i\langle\{\phi(x_1),\phi(x_2)\}\rangle\, ,
 \label{drr}\\
 \Delta_{ra}(x_1, x_2) &=&\Delta_{11}(x_1, x_2)-\Delta_{12}(x_1, x_2)
 \nonumber\\
   &=&-i\theta(x_1^0-x_2^0)\langle [\phi(x_1),\phi(x_2)]\rangle\, ,
 \label{dra}\\
 \Delta_{ar}(x_1, x_2) &=&\Delta_{11}(x_1, x_2)-\Delta_{21}(x_1, x_2)
   \nonumber\\
   &=&-i\theta(x_2^0-x_1^0)\langle [\phi(x_2),\phi(x_1)]\rangle\, ,
  \label{dar}\\
 \Delta_{aa}(x_1, x_2) &=& 0\, .
 \label{daa}
 \end{eqnarray}
 \end{mathletters}
Obviously $\Delta_{ra}(x_1, x_2)$ and $\Delta_{ar}(x_1,x_2)$ are the 
usual retarded and advanced linear response functions. In thermal 
equilibrium the correlation function $\Delta_{rr}(x_1, x_2)$ and 
the linear response functions satisfy the Fluctuation-Dis\-si\-pation
Theorem \cite{Callen}
 \begin{equation}
 \label{FDT1}
 \Delta_{rr}(k)=\bigl(1{+}2n(k^0)\bigr)
 \bigl(\Delta_{ra}(k)-\Delta_{ar}(k)\bigr)
 \end{equation}
in momentum space where $n(k^0){\,=\,}1/(e^{\beta k^0}{-}1)$ is the 
Bose distribution. Eqs.~(\ref{d}) are inverted by
 \begin{equation}
 \label{Gtrans}
   \Delta_{a_1 a_2}(x_1,x_2) = \Delta_{\alpha_1\alpha_2}(x_1,x_2)
     Q_{\alpha_1 a_1}Q_{\alpha_2 a_2}\, ,
 \end{equation}
where repeated indices are summed over and
 \begin{equation}
 \label{Q}
    Q_{a 1}=-Q_{a 2}=Q_{r 1}=Q_{r 2}={1\over \sqrt{2}}
 \end{equation}
are the four elements of the orthogonal Keldysh transformation for
2-point functions \cite{Keldysh}. For example, one can express in 
momentum space the (12)-component of the 2-point function related to 
the Wightman function in Eq.~(\ref{eta1}) as
 \begin{equation}
 \label{G12}
  \Delta^{(\pi\pi)}_{12}(k)={1\over 2}
  \Bigl(\Delta^{(\pi\pi)}_{rr}(k)-\Delta^{(\pi\pi)}_{ra}(k)
   +\Delta^{(\pi\pi)}_{ar}(k)\Bigr) .
 \end{equation}
Using the Fluctuation-Dissipation Theorem and taking into account
$\Delta^*_{ra}(k){\,=\,}\Delta_{ar}(k)$, we see that the 
(12)-com\-po\-nent of the 2-point function $\Delta^{(\pi\pi)}$ for 
the composite field $\pi_{ij}$ is purely imaginary in momentum space.
Eq.~(\ref{eta1}) can thus be expressed as
 \begin{equation}
 \label{eta2}
   \eta = {{i\beta}\over 20}\lim_{p^0,\bbox{p} \to 0}
          \Delta^{(\pi\pi)}_{12}(p)
   = -{{\beta}\over 20}\lim_{p^0,\bbox{p} \to 0}
         {\rm Im\,} \Delta^{(\pi\pi)}_{12}(p)\,.
 \end{equation}

 \begin{figure}
 \epsfxsize 80mm \epsfbox{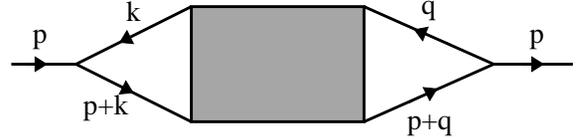}
 \vskip 0.4cm
 \caption{\label{F1}
  \small The shear viscosity can be expressed as an integral over
  the 4-point single particle Green function, with two pairs of external
  legs joined to a point, in the limit of zero external momentum $p$.}
 \end{figure}

Since each composite field $\pi_{ij}$ is made from two single-particle
fields $\phi$, the shear viscosity $\eta$ is related to the 4-point
Green function of the $\phi$ field, as illustrated in Fig.~\ref{F1}.
Substituting $\pi_{ij}$ into Eq.~(\ref{eta2}) one gets \cite{Jeon1,Jeon2}
 \begin{eqnarray}
 \label{eta3}
   \eta &=& {{\beta}\over 5}\lim_{p^0,\bbox{p} \to 0}
          \int{d^4k\over (2\pi)^4}\int{d^4q\over (2\pi)^4}
          J_{ij}(-k, p{+}k)
  \nonumber\\
    &\times& {\rm Im\,}G_{1122}(-k, p{+}k, q, -p{-}q)
      J^{ij}(q, -p{-}q)\, ,
 \end{eqnarray}
where the function $J_{ij}(p,q)$ joins two $\phi$ propagators to a point 
(see Fig.~\ref{F1}),
 \begin{equation}
 \label{join}
   J_{ij}(p,q) = p_i q_j-{\textstyle{1\over 3}}\delta_{ij}\,
   \bbox{p}\cdot\bbox{q}\,,
 \end{equation}
and $G_{1122}(-k, p{+}k, q, -p{-}q)$ in Eq.(\ref{eta3}) denotes
the Fourier-transformed $(1122)$-component of the $\phi$ 4-point Green
function. Its 16 CTP components are defined as
 \begin{eqnarray}
 \label{Ga1a2a3a4}
   &&G_{a_1 a_2 a_3 a_4}(x_1,x_2,x_3,x_4)
 \\
   &&\quad\equiv (-i)^3
   \left\langle T_p\left[\phi_{a_1}(x_1)\phi_{a_2}(x_2)
   \phi_{a_3}(x_3)\phi_{a_4}(x_4)\right]\right\rangle
 \nonumber\\
   &&\quad={\textstyle{1\over 2}}
     G_{{\alpha}_1 {\alpha}_2 {\alpha}_3 {\alpha}_4}(x_1,x_2,x_3,x_4)
     Q_{\alpha_1 a_1}Q_{\alpha_2 a_2}Q_{\alpha_3 a_3} Q_{\alpha_4 a_4}\,.
 \nonumber
 \end{eqnarray}
where $G_{{\alpha}_1 {\alpha}_2 {\alpha}_3 {\alpha}_4}(x_1,x_2,x_3,x_4)$ 
is the corresponding 4-point Green function in the $(r,a)$ basis,
defined as \cite{Chou,Wang3}
 \begin{eqnarray}
 \label{Gaf1af2af3af4}
   &&G_{{\alpha}_1 {\alpha}_2 {\alpha}_3 {\alpha}_4}(x_1,x_2,x_3,x_4)\equiv
   (-i)^3 2^{n_r-1}
 \nonumber\\
   &&\quad\times
   \left\langle T_p\left[\phi_{{\alpha}_1}(x_1)\phi_{{\alpha}_2}(x_2)
   \phi_{{\alpha}_3}(x_3)\phi_{{\alpha}_4}(x_4)\right]
   \right\rangle\, .
 \end{eqnarray}
Generally, in the $(r,a)$ basis the $n$-point Green function with only 
$a$ indices vanishes: $G_{aa\dots a}{\,=\,}0$ \cite{Wang3}.

\subsection{The need for ladder resummation}
\label{sec2b}

In the one-loop approximation the 4-point function of Fig.~\ref{F1} 
reduces to a product of two 2-point Green functions $\Delta_{12}$ 
which, in the limit of vanishing external momentum, have opposite
momenta. The (12)-component of the 2-point function can be 
expressed in terms of its spectral function $\rho$ \cite{Wang1}.
For noninteracting particles, the poles of the two spectral functions 
with opposite loop momenta pinch the real axis in the complex energy 
plane from above and below, resulting in a pinch singularity for the
integral over the energy circulating in the loop \cite{fn1}. This 
singularity can be cured by using resummed $\phi$ propagators with a 
two-loop self-energy in the denominator \cite{Jeon2,Parwani,Wang4}. 
This resummation generates scalar quasi-particles with a finite thermal 
(collisional) width and lifetime \cite{Wang4} which shifts the pinching 
poles away from the real energy axis.

\begin{figure}
 \epsfxsize 80mm \epsfbox{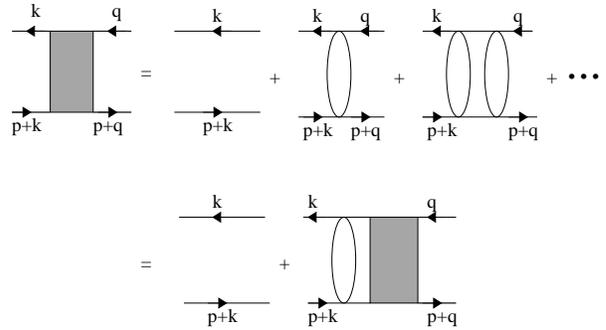}
 \vskip 0.4cm
 \caption{\label{F2}
  \small In $\lambda\phi^4$ theory the summation of an infinite number of
  planar ladder diagrams contributing to the shear viscosity at leading 
  order results in a Bethe-Salpeter equation for the 4-point Green
  function.}
 \end{figure}

As shown by Jeon \cite{Jeon1,Jeon2}, however, this regularization of 
the energy loop integral is not enough to produce a reliable 
leading-order result for the shear viscosity. We rephrase his arguments, 
using weakly coupled $\lambda\phi^4$ theory as an example. Using the 
two-loop resummed propagator in the basic one-loop skeleton diagram
for $\eta$, the propagator pair sharing the same loop momentum 
contributes a factor ${\cal O}(1/\lambda^2)$ to the shear viscosity. 
(We call this the ``nearly pinching poles'' contribution.) Clearly 
such terms involving ``nearly pinching poles'' are dominant in 
the weak coupling approximation. For multi-loop ladder diagrams with 
parallel rungs formed by one-loop diagrams connecting the two ``side 
rail'' propagators (see Fig.~\ref{F2}), each pair of side rails 
shares the same loop momentum as the external momentum $p$ approaches 
zero; correspondingly, its frequency integral generates a ``nearly 
pinching poles'' contribution $\sim 1/\lambda^2$ which just compensates 
the factor $\lambda^2$ from the two extra vertices. 

All multi-loop ladder diagrams with parallel rungs thus contribute at 
the same order as the simple one-loop diagram. On the other hand, 
multi-loop ladders with crossing rungs have different 
momenta on some of their side rails; since the integrals over the 
corresponding loop energies are free from ``nearly pinching poles'' 
contributions, these diagrams are genuinely suppressed by additional 
powers of $\lambda^2$. Similar arguments rule out leading-order 
contributions from all other multi-loop diagrams \cite{Jeon2}. At 
leading order we must therefore only resum the infinite ladders with 
parallel rungs. In the CTP formalism this is achieved by solving a 
Bethe-Salpeter equation for the 4-point Green function.

In the Keldysh basis, the BS integral equation couples the component
$G_{1122}$ needed in Eq.~(\ref{eta3}) to three other components of the
4-point function: $G_{1222}$, $G_{2122}$, and $G_{2222}$. The key 
technical issue is therefore whether and how the BS equation can be 
decoupled. In the following we will show that in the weak coupling 
limit this is indeed possible, and that this is most conveniently 
achieved in the $(r,a)$ basis: all components of the bare 4-point 
vertex involving an even number of $a$ indices vanish (see 
Eq.~(\ref{coupling})), and the propagator $\Delta_{aa}{\,=\,}0$ (see 
Eq.~(\ref{daa})). When decomposing $G_{1122}$ into a linear combination 
of the sixteen $(r,a)$ 4-point functions an important simplification 
arises from the generalized FTD \cite{Wang3} (see also \cite{CHS} and 
recent work by Guerin\cite{Guerin}) which relates these functions among 
each other, reducing the number of independent 4-point vertices to seven. 
We will see that, at leading order in the weak coupling approximation, 
only one of these seven independent functions contributes to the shear 
viscosity and its BS equation decouples. Again, the generalized FDT 
plays an important role in the decoupling procedure.

\subsection{Contributions to $\eta$ in the $(r,a)$ basis}
\label{sec2c}

Let us now follow the steps of expressing Eq.~(\ref{eta3}) in $(r,a)$
components. From the KMS condition \cite{Kubo} one derives in momentum 
space \cite{Chou,Wang3}
 \begin{eqnarray}
 \label{G*}
    &&G^*_{a_1 a_2 a_3 a_4}(k_1, k_2,k_3,k_4)
  \nonumber\\
    &&\qquad = -G_{{\bar a}_1 {\bar a}_2 {\bar a}_3 {\bar a}_4}
      (k_1, k_2, k_3, k_4)
      \prod_{\{i|a_i=2\}} \!\!\! e^{\beta k^0_i}\, ,
  \end{eqnarray}
where all momenta flow into the vertex such that 
$k_1{+}k_2{+}k_3{+}k_4{\,=\,}0$, the star denotes complex conjugation, 
and ${\bar a}_i{\,=\,}2,1$ for $a_i{\,=\,}1,2$, respectively. We can 
thus reexpress $G_{1122}(-k,p{+}k,q,-p{-}q)$ as
 \begin{eqnarray}
 \label{G1122a}
   &&G_{1122}(-k,p{+}k,q,-p{-}q)
 \nonumber\\
   &&\quad= -n(p^0)\Bigl(G_{1122} + G^*_{2211}\Bigr)(-k,p{+}k,q,-p{-}q).
 \end{eqnarray}
We combine this with Eq.~(\ref{Ga1a2a3a4}) to decompose ${\rm Im\,}G_{1122}$ 
in the $(r,a)$ basis:
 \begin{eqnarray}
 \label{G1122b}
   {\rm Im\,}&&G_{1122}(-k,p{+}k,q,-p{-}q)
 \nonumber\\
   &&=-n(p^0){\rm Im\,}\Bigl(G_{1122}
            +G^*_{2211}\Bigr)(-k,p{+}k,q,{-}p{-}q)
 \nonumber\\
   &&={\textstyle{1\over 4}}n(p^0){\rm Im\,}
      \Bigl(G_{aaar}{+}G_{aara}{-}G_{araa}{-}G_{raaa}{+}G_{rrra}
 \nonumber\\
   &&\quad {+}G_{rrar}{-}G_{rarr}{-}G_{arrr}\Bigr)(-k,p{+}k,q,-p{-}q)\,.
 \end{eqnarray}
The first four terms on the right hand side involve the fully retarded 
Green functions defined in \cite{Lehmann}, $G_{raaa}$, $G_{araa}$, 
$G_{aara}$, and $G_{aaar}$. They are part of the set of 7 independent
4-point functions chosen in Ref.~\cite{Wang3} which additionally
includes $G_{arra}$, $G_{arar}$, and $G_{aarr}$. The last four terms 
in Eq.~(\ref{G1122b}) are not part of this set, but can be expressed
through these 7 functions via the generalized FDT \cite{Wang3}:
\end{multicols}
 \begin{mathletters}
 \label{G4sol}
 \begin{eqnarray}
   G_{rrra}(k_1,k_2,k_3,k_4) =&&\Bigl[N_2 N_3G_{raaa}
      -N_2 N_4 N^{(13)}_{(24)}G^*_{araa}
      -N_3 N_4 N^{(12)}_{(34)}G^*_{aara}
      -\Bigl( N^{(13)}_{(24)}+N_3^2 N^{(12)}_{(34)}\Bigr) G^*_{aaar}
   \nonumber\\
    &&-N_1 G_{arra}-N_2 N^{(13)}_{(24)}G^*_{arar}
      -N_3 N^{(12)}_{(34)}G^*_{aarr}\Bigr](k_1,k_2,k_3,k_4)\, ,
  \label{G4sol2}\\
   G_{rrar}(k_1,k_2,k_3,k_4) =&&\Bigl[N_2 N_4G_{raaa}
      -N_2 N_3 N^{(14)}_{(23)}G^*_{araa}
      -\Bigl( N^{(12)}_{(34)}+N_2^2 N^{(14)}_{(23)}\Bigr) G^*_{aara}
      -N_3 N_4 N^{(12)}_{(34)}G^*_{aaar}
   \nonumber\\
    &&-N_2 N^{(14)}_{(23)}G^*_{arra}-N_1 G_{arar}
      -N_4 N^{(12)}_{(34)}G^*_{aarr}\Bigr](k_1,k_2,k_3,k_4)\, ,
  \label{G4sol3}\\
   G_{rarr}(k_1,k_2,k_3,k_4) =&&\Bigl[N_3 N_4G_{raaa}
      -\Bigl( N^{(14)}_{(23)}+N_4^2 N^{(13)}_{(24)}\Bigr) G^*_{araa}
      -N_2 N_3 N^{(14)}_{(23)}G^*_{aara}-N_2 N_4 N^{(13)}_{(24)}G^*_{aaar}
   \nonumber\\
    &&-N_3 N^{(14)}_{(23)}G^*_{arra}
      -N_4 N^{(13)}_{(24)}G^*_{arar}-N_1 G_{aarr}\Bigr](k_1,k_2,k_3,k_4)\, ,
  \label{G4sol4}\\
   G_{arrr}(k_1,k_2,k_3,k_4) =&&\Bigl[\Bigl( 1+ N_2 N_3+ N_2 N_4
      + N_3 N_4\Bigr)G^*_{raaa}
      -N_3 N_4 G_{araa}-N_2 N_4 G_{aara}-N_2 N_3 G_{aaar}
   \nonumber\\
    &&-N_4 G_{arra}-N_3 G_{arar}-N_2
      G_{aarr}\Bigr](k_1,k_2,k_3,k_4)\, .
 \end{eqnarray}
 \end{mathletters}
\begin{multicols}{2}
\noindent
Here $N_i{\,=\,}N(k_i^0){\,=\,}1{+}2n(k_i^0)$ and 
$N^{(ij)}_{(lm)}{\,=\,}{N_i{+}N_j\over N_l{+}N_m}$,
with $k^0_1{+}k^0_2{+}k^0_3{+}k^0_4{\,=\,}0$. Substituting 
Eqs.~(\ref{G4sol}) into Eq.~(\ref{G1122b}) we express 
${\rm Im\,}G_{1122}(-k,p{+}k,q,-p{-}q)$ as
 \begin{eqnarray}
 \label{imaginary}
  &&{\rm Im\,}G_{1122}(-k,p{+}k,q,-p{-}q)
 \nonumber\\
  &&={\rm Im\,}\bigl[{\cal A} G_{aarr}+ {\cal B} G_{arar}+ {\cal C} G_{arra}
  + {\cal D} G_{raaa} \\
  &&\quad\quad\ + {\cal E} G_{araa} + {\cal F} G_{aara}+ {\cal G} G_{aaar}
    \bigr](-k,p{+}k,q,-p{-}q)\, ,
 \nonumber
 \end{eqnarray}
where the coefficients ${\cal A},{\cal B},\cdots,{\cal G}$ involve
combinations of thermal distribution functions. In the relevant limit 
$p^0{\,\to\,}0$ they are given, up to terms of order $\beta p^0$, by
 \begin{mathletters}
 \label{coeff}
 \begin{eqnarray}
  {\cal A} &\approx& -{{\beta p^0 n(p^0)}\over 4} (N^2_k{-}1)\,,       
 \label{coeff7}\\
  {\cal B} &\approx& {\cal C} \approx 
  {\beta p^0 n(p^0)\over 8} \Bigl( (N^2_k{-}1)
  + (N^2_q{-}1)\Bigr),
 \label{coeff6}\\
  {\cal D} &\approx& {\cal F} + {\cal G}\,,
 \label{coeff1}\\
  {\cal E} &\approx& 0\,,
 \label{coeff2}\\
  {\cal F} &\approx& {\beta p^0 n(p^0)\over 8} 
   \Bigl( N_k (N^2_q{-}1) + N_q (N^2_k{-}1)\Bigr),
 \label{coeff3}\\
  {\cal G} &\approx& {\beta p^0 n(p^0)\over 8} 
   \Bigl( N_k (N^2_q{-}1) - N_q (N^2_k{-}1)\Bigr).
 \label{coeff4}
 \end{eqnarray}
 \end{mathletters}
Here $N_k{\,=\,}N(k^0){\,=\,}1{+}2 n(k^0)$ and similarly for $N_q$. 
In the zero external momentum limit the explicitly shown terms
are finite since $\lim_{p^0\to 0} \beta p^0\, n(p^0)=1$, whereas
the dropped terms of order $\beta p^0$ and higher vanish. Since
${\cal E}{\,=\,0}$ in this limit, $G_{araa}$ does not contribute to 
the shear viscosity.

The $n$-point functions involving only elementary fields $\phi$ are 
symmetric under particle exchange:
 \begin{eqnarray}
 \label{Gsym}
   &&G_{\dots\alpha_i\dots\alpha_j\dots}
    (\dots,k_i,\dots,k_j,\dots)
 \nonumber\\
   &&\quad =G_{\dots\alpha_j\dots\alpha_i\dots}
    (\dots,k_j,\dots,k_i,\dots)\,.
 \end{eqnarray}
Moreover, the factor $J_{ij}(-k,k)J^{ij}(q,-q)$ in Eq.~(\ref{eta3}) 
is invariant under each of the following changes of integration 
variables: $k{\,\leftrightarrow\,}q$, $k{\,\leftrightarrow\,}{-}k$ and 
$q{\,\leftrightarrow\,}{-}q$. Using Eq.~(\ref{Gsym}), both 
$G_{raaa}({-}k,k,q,{-}q)$ under the variable change 
$k{\,\leftrightarrow\,}q$ and $G_{aara}({-}k,k,q,{-}q)$ under the 
variable change $q{\,\rightarrow\,}{-}q$ become equal to 
$G_{aaar}({-}k,k,q,{-}q)$. Performing the same variable changes on 
the corresponding prefactors ${\cal D}$ and ${\cal F}$ we can 
substitute in Eq.~(\ref{eta3}) 
 \begin{eqnarray}
 \label{Gsimple}
 [{\cal D} G_{raaa} &+& {\cal F} G_{aara} + {\cal G} G_{aaar}](-k, k, q, -q)
 \nonumber\\
 &\to& {\textstyle{1\over 4}} N_k (N_q^2{-}1)\, G_{aaar}(-k, k, q, -q).
 \end{eqnarray} 
Using Eq.~(\ref{Gsym}) together with $N({-}k^0){\,=\,}{-}N(k^0)$
this is seen to be an odd function of $k$ which integrates to zero
in Eq.~(\ref{eta3}). Consequently, {\em none of the four fully retarded 
4-point functions contributes to the shear viscosity}.

Similarly, under the variable change $q{\,\rightarrow\,}{-}q$ we 
have $G_{arar}({-}k,k,q,{-}q){\,\to\,}G_{arra}({-}k,k,q,{-}q)$ such
that 
 \begin{eqnarray}
 \label{Gsimple2}
  &&[{\cal B} G_{arar} + {\cal C} G_{arra}](-k,k,q,-q)
 \nonumber\\
  &&\qquad \to {\textstyle{1\over4}} \Bigl((N_k^2{-}1)+(N_q^2{-}1)\Bigr)
    G_{arra}(-k,k,q,-q)
 \nonumber\\
  && \qquad \to {\textstyle{1\over 2}}(N_k^2{-}1)G_{arra}(-k,k,q,-q)\,,
 \end{eqnarray} 
where the second substitution involves the variable change 
$k{\,\leftrightarrow\,}q$. We will see below that in the
weak coupling limit this term does not contribute to $\eta$ at 
leading order either, leaving only the contribution from 
$G_{aarr}$.

\subsection{Bethe-Salpeter equation for 
            $G_{\alpha_1\alpha_2\alpha_3\alpha_4}$}
\label{sec2d}

Referring to Fig.~\ref{F2} above and Fig.~\ref{F5} further below, the 
BS equation for the 4-point Green function 
$G_{\alpha_1\alpha_2\alpha_3\alpha_4}$ can be expressed in the $(r,a)$ 
basis as 
 \begin{eqnarray}
 \label{Galpha}
   &&i^3 G_{\alpha_1\alpha_2\alpha_3\alpha_4}(-k,k,q,-q)=
 \nonumber\\
   && \left[i\Delta_{\alpha_1\alpha_3}(-k)\right]
      \left[i\Delta_{\alpha_2\alpha_4}(k)\right]
      (2\pi)^4\delta^4(k{-}q)+
 \nonumber\\
    &&\left[i\Delta_{\alpha_1\beta_1}(-k)\right]
    \left[i\Delta_{\alpha_2\gamma_1}(k)\right]
    \int{d^4l\over (2\pi)^4}
    K_{\beta_1\gamma_1\beta_4\gamma_4}(-k,k,l,-l)
 \nonumber\\
    && \ \times
       \left[i^3
       G_{\beta_4\gamma_4\alpha_3\alpha_4}(-l,l,q,-q)\right]\,.
 \end{eqnarray}
As always repeated indices are summed over, and
$K_{\beta_1\gamma_1\beta_4\gamma_4}$ is the kernel (an amputed 1PI 
4-point vertex function) of the integral equation. For the component
$G_{arra}$ we have
 \begin{eqnarray}
 \label{Garra}
   &&G_{arra}(-k,k,q,-q)
 \nonumber\\
   &&\quad = -i\Delta_{ra}(k)\Delta_{ra}(k)
      (2\pi)^4\delta^4(k{-}q)
 \nonumber\\
    &&\quad -\Delta_{ra}(k)\Delta_{rr}(k)
    \int{d^4l\over (2\pi)^4}
    K_{rr\beta_4\gamma_4}(-k,k,l,-l)
 \nonumber\\
    &&\qquad\times G_{\beta_4\gamma_4 ra}(-l,l,q,-q)
 \nonumber\\
    &&\quad -\Delta_{ra}(k)\Delta_{ra}(k)
    \int{d^4l\over (2\pi)^4}
    K_{ra\beta_4\gamma_4}(-k,k,l,-l)
 \nonumber\\
    &&\qquad\times G_{\beta_4\gamma_4 ra}(-l,l,q,-q)\, .
 \end{eqnarray}
Here we used 
$\Delta_{\alpha_1\alpha_2}({-}k){\,=\,}\Delta_{\alpha_2\alpha_1}(k)$ and
$\Delta_{aa}(k){\,=\,}0$. In the second term we can reexpress $\Delta_{rr}$
in terms of $\Delta_{ra}$ and $\Delta_{ar}$ using the FDT (\ref{FDT1}).
Since $\Delta_{ra}(k)$ and $\Delta_{ar}(k)$ are retarded and advanced 
Green functions, respectively, the poles of $\Delta_{ra}(k)\Delta_{ra}(k)$ 
and $\Delta_{ar}(k)\Delta_{ar}(k)$ in the complex $k^0$ plane lie on 
the same side of the real $k^0$ axis. Hence, in the limit $\lambda{\,\to\,}0$, 
they do not generate a pinch singularity in the integrand of Eq.~(\ref{eta3}).
The term $\sim \Delta_{ra}(k)\Delta_{ar}(k){\,=\,}| \Delta_{ra}(k)|^2$, 
on the other hand, does generate a pinch singularity as $\lambda{\,\to\,}0$; 
in the weak coupling limit, this term will thus dominate in Eq.~(\ref{eta3}) 
over the two other ones. With this ``nearly pinching poles'' approximation 
we can rewrite Eq.~(\ref{Garra}) as
 \begin{eqnarray}
 \label{Garra+}
   &&G_{arra}(-k,k,q,-q) = N(k^0)|\Delta_{ra}(k)|^2
 \\
   &&\quad\times
   \int{d^4l\over (2\pi)^4} K_{rr\beta_4\gamma_4}(-k,k,l,-l)\,
   G_{\beta_4\gamma_4 ra}(-l,l,q,-q)\, .
 \nonumber
 \end{eqnarray}
Using $N({-}k^0){\,=\,}{-}N(k^0)$ and the analogue of the symmetry 
relations (\ref{Gsym}) for the amputated 1PI vertex \cite{Wang3},
we see that ${1\over 2}(N_k^2{-}1)G_{arra}(-k,k,q,-q)$ is an odd 
function of $k$ and thus gives a vanishing leading order contribution 
to the shear viscosity in Eq.~(\ref{eta3}). This leaves only $G_{aarr}$:
 \begin{eqnarray}
 \label{eta4}
   \eta &=& -{\beta\over 5}\int{d^4k\over (2\pi)^4}
            n(k^0)\bigl(1{+}n(k^0)\bigr)I_{\pi,lm}(k)
 \nonumber\\
   &&\times\int{d^4q\over (2\pi)^4}
     {\rm Im\, }G_{aarr}(-k,k,q,-q)I_\pi^{lm}(q)\,.
 \end{eqnarray}
Here $I_{\pi,lm}(k)\equiv -J_{lm}(-k,k)$ is the notation used in \cite{Jeon2}.

\subsection{Decoupling the BS equation for $G_{aarr}$}
\label{sec2e}

We now show that the BS equation for $G_{aarr}$ decouples from the 
other components of $G_{\alpha_1\alpha_2\alpha_3\alpha_4}$ in the 
``nearly pinching poles'' approximation. From Eq.~(\ref{Galpha}) 
we obtain
 \begin{eqnarray}
 \label{Gaarr}
   &&G_{aarr}(-k,k,q,-q)
 \nonumber\\
   &&\quad = -\Delta_{ra}(k)\Delta_{ar}(k)\Bigl[
      i(2\pi)^4\delta^4(k{-}q)
 \nonumber\\
    &&\quad +\int{d^4l\over (2\pi)^4}
    K_{rr\beta_4\gamma_4}(-k,k,l,-l)
    G_{\beta_4\gamma_4 rr}(-l,l,q,-q)\Bigr]
 \nonumber\\
   &&\quad = -\Delta_{ra}(k)\Delta_{ar}(k)\Bigl\{
      i(2\pi)^4\delta^4(k{-}q)
 \nonumber\\
    &&\quad+\int{d^4l\over (2\pi)^4}\Bigl[
      K_{rrra}(-k,k,l,-l)G_{rarr}(-l,l,q,-q)
 \nonumber\\
    &&\qquad\qquad +K_{rrar}(-k,k,l,-l)G_{arrr}(-l,l,q,-q)
 \nonumber\\
    &&\qquad\qquad +K_{rraa}(-k,k,l,-l)G_{aarr}(-l,l,q,-q)\Bigr]\Bigr\}\,.
 \end{eqnarray}
For the second equality we used $K_{rrrr}{\,=\,}0$ which is the analogue 
for the amputated 1PI 4-point vertex of the previously mentioned relation 
$G_{aaaa}{\,=\,}0$ \cite{Wang3}. At this point $G_{aarr}$ still couples 
to $G_{rarr}$ and $G_{arrr}$. For later convenience we introduce a 
function $M$ which is obtained by truncating two of the external legs 
of the 4-point Green function $G$:
 \begin{eqnarray}
 \label{M}
  &&G_{\alpha_1\alpha_2\alpha_3\alpha_4}(-l,l,q,-q) =
 \nonumber\\
  && \left[i\Delta_{\alpha_1\beta_1}(-l)\right]
    \left[i\Delta_{\alpha_2\beta_2}(l)\right]
    M_{\beta_1\beta_2\alpha_3\alpha_4}(-l,l,q,-q)\, .
 \end{eqnarray}
It follows that
 \begin{eqnarray}
 \label{GaarrM}
  G_{aarr}&&(-l,l,q,-q)
 \nonumber\\
  &&= -\Delta_{ra}(l)\Delta_{ar}(l) M_{rrrr}(-l,l,q,-q)\,.
 \end{eqnarray}
As above we now use the ``nearly pinching poles'' approximation, ignoring 
terms involving $\Delta_{ra}(l)\Delta_{ra}(l)$ and 
$\Delta_{ar}(l)\Delta_{ar}(l)$ relative to terms involving
$\Delta_{ra}(l)\Delta_{ar}(l)$. Using Eqs.~(\ref{daa}) and (\ref{FDT1}),
we can then write
 \begin{mathletters}
 \label{GM}
 \begin{eqnarray}
  G_{rarr}&&(-l,l,q,-q)
 \nonumber\\
  &&= -N(l^0)\Delta_{ra}(l)\Delta_{ar}(l)
    M_{rrrr}(-l,l,q,-q)
 \nonumber\\
   &&= N(l^0)G_{aarr}(-l,l,q,-q)\, ,
 \label{GrarrM}\\
  G_{arrr}&&(-l,l,q,-q)
 \nonumber\\
  &&= N(l^0)\Delta_{ra}(l)\Delta_{ar}(l)
    M_{rrrr}(-l,l,q,-q)
 \nonumber\\
   &&= -N(l^0)G_{aarr}(-l,l,q,-q)\, .
 \label{GarrrM}
 \end{eqnarray}
 \end{mathletters}
These leading order relations decouple the BS equation (\ref{Gaarr})
for $G_{aarr}$; we find
 \begin{eqnarray}
 \label{Gaarr+}
   &&G_{aarr}(-k,k,q,-q)
 \nonumber\\
   &&\quad = -\Delta_{ra}(k)\Delta_{ar}(k)\Bigl[
      i(2\pi)^4\delta^4(k{-}q)
 \nonumber\\
    &&\quad+\int{d^4l\over (2\pi)^4}\,
      {\cal K}(-k,k,l,-l) \, G_{aarr}(-l,l,q,-q)\Bigr]
 \end{eqnarray}
with the kernel 
 \begin{eqnarray}
 \label{kernel}
 {\cal K}&&(-k,k,l -l)=
 \nonumber\\
  &&\Bigl(N(l^0)\left(K_{rrra}{-}K_{rrar}\right)
     +K_{rrra}\Bigr)(-k,k,l,-l)\,.
 \end{eqnarray}

\subsection{An integral equation for the viscosity}
\label{sec2f}

For further simplification we introduce the 2-point spectral density
 \begin{equation}
 \label{density}
    \rho(k)=i[\Delta_{ra}(k)-\Delta_{ar}(k)]\, .
 \end{equation}
Writing the full retarded propagator $\Delta_{ra}(k)$ as
 \begin{equation}
 \label{fullret}
  \Delta_{ra}(k)={1\over {k^2-{\rm Re\, }\Sigma(k)
           +i\,{\rm Im\, }\Sigma(k)}}
 \end{equation}
and using $\Delta_{ar}(k){\,=\,}\Delta^*_{ra}(k)$ we deduce
 \begin{eqnarray}
 \label{density+}
  &&\rho(k)={{2\,{\rm Im\,}\Sigma(k)}\over 
             {\bigl(k^2-{\rm Re\,}\Sigma(k)\bigr)^2 +
              \bigl({\rm Im\, }\Sigma(k)\bigr)^2}}\,,
 \\
 \label{pinch}
  &&\Delta_{ra}(k)\Delta_{ar}(k)= |\Delta_{ra}(k)|^2=
    {\rho(k)\over {2{\rm Im\,}\Sigma(k)}}\,.
 \end{eqnarray}
Substituting Eq.~(\ref{pinch}) into Eq.~(\ref{GaarrM}) we get
 \begin{eqnarray}
 \label{ImGaarr}
   {\rm Im\,}G_{aarr}&&(-k,k,q,-q) = 
 \nonumber\\
 &&-{\rho(k)\over {2{\rm Im\, }\Sigma(k)}}
   {\rm Im\, }M_{rrrr}(-k,k,q,-q)\,,
 \end{eqnarray}
and the decoupled BS equation (\ref{Gaarr+}) can be written as
 \begin{eqnarray}
 \label{Mrrrr}
    &&{\rm Im\, }M_{rrrr}(-k,k,q,-q)=(2\pi)^4\delta^4(k{-}q)
 \\
    &&+\int{d^4l\over (2\pi)^4}{\rho(l)\over {2{\rm Im\, }\Sigma(l)}}
    {\cal K}(-k,k,l,-l)
    {\rm Im\, }M_{rrrr}(-l,l,q,-q) .
 \nonumber
 \end{eqnarray}
Defining further
 \begin{equation}
 \label{Dpi}
   D_{\pi}^{lm}(k) = \int{d^4q\over (2\pi)^4}\,I_{\pi}^{lm}(q)\,
                     {\rm Im\, }M_{rrrr}(-k,k,q,-q)
 \end{equation}
and using it in Eqs.~(\ref{eta4}) and (\ref{Mrrrr}) we finally arrive at 
 \begin{eqnarray}
 \label{eta5}
   \eta = {\beta\over 10}{\int}{d^4k\over (2\pi)^4}\,
          n(k^0)\bigl(1{+}n(k^0)\bigr)
         {\rho(k)\, I_\pi(k)\cdot D_\pi(k)\over {\rm Im\, }\Sigma(k)}.
 \end{eqnarray}
Here $I_\pi\cdot D_\pi \equiv I_{\pi,lm}D_\pi^{lm}$, and $D_{\pi}(k)$ 
satisfies the integral equation (suppressing the tensor indices on
$I_\pi,D_\pi$)
 \begin{eqnarray}
 \label{Dpion}
    D_{\pi}(k)=I_{\pi}(k)-{\int}{d^4l\over (2\pi)^4}
    {\cal K}(-k,k,l,-l)\,
    {\rho(l)\,D_{\pi}(l)\over {2\,{\rm Im\, }\Sigma(l)}}.
 \end{eqnarray}
These last two expressions were previously obtained in \cite{Jeon2,Wang2}.
The present derivation, however, does not use a specific form of the
interaction Lagrangian; within the ``nearly pinching poles'' approximation,
Eqs.~(\ref{eta5}) and (\ref{Dpion}) form a general result for {\em all} 
scalar field theories, with different forms of the interaction Lagrangian 
resulting in different kernels ${\cal K}(-k,k,l,-l)$. We expect a similar 
integral equation to hold for the leading order viscosity in pure 
(quarkless) QCD, with modified expressions for $I_\pi$ and ${\cal K}$. 
In the following Section, we evaluate ${\cal K}$ in the CTP formalism 
for the $\lambda\phi^4$ and $g\phi^3{+}\lambda\phi^4$ theories and show 
that the results agree with those obtained by Jeon \cite{Jeon2} in the 
imaginary time formalism.

\section{The shear viscosity kernel in scalar field theories}
\label{sec3}
\subsection{Massless $\lambda\phi^4$ theory}
\label{sec3a}

The Lagrangian for massless $\phi^4$ theory reads
 \begin{equation}
 \label{lag0}
    {\cal L}_0 = {1\over 2} (\partial_{\mu}\phi)^2 
                 - {\lambda \over 4!} \phi^4,
 \end{equation}
and we consider it in the weak coupling limit $\lambda\ll 1$. For this 
theory the kernel of the integral equation (see Fig.~\ref{F2}) can be 
expressed as
 \begin{eqnarray}
 \label{ker}
   K_{\beta_1\gamma_1\beta_4\gamma_4}&&(-k,k,l,-l)=
   {1\over 2}{\int}{d^4s\over (2\pi)^4}
    \lambda_{\gamma_1\gamma_2\gamma_3\gamma_4}
    \lambda_{\beta_1\beta_2\beta_3\beta_4}
 \nonumber\\
   &&\times\Delta_{\gamma_2\beta_2}(s)\Delta_{\gamma_3\beta_3}(s+l-k)\, ,
 \end{eqnarray}
where the prefactor ${1\over2}$ is the symmetry factor associated with
the bubble connecting the two lines, and the bare 4-point vertex
in the $(r,a)$ basis is given by \cite{Wang3}
 \begin{equation}
 \label{coupling}
  \lambda_{\alpha_1\alpha_2\alpha_3\alpha_4}=
  {\lambda\over 4}\Bigl(1-(-1)^{n_a}\Bigr)\, ,
 \end{equation}
$n_a$ being the number of $a$ indices among $(\alpha_1,\alpha_2,
\alpha_3,\alpha_4)$. The last equation implies that all bare vertices 
must involve an odd number of $a$ indices (in particular 
$\lambda_{rrrr}{\,=\,}0$).

Using Eq.~(\ref{coupling}) and $\Delta_{aa}(k)=0$, Eq.~(\ref{ker}) gives
 \begin{mathletters}
 \label{kphi4}
 \begin{eqnarray}
 \label{krrra}
 K&&_{rrra}(-k,k,l,-l) = {\lambda^2\over 8}{\int}{d^4s\over (2\pi)^4}
 \\
  &&\quad \times\Bigl[\Delta_{rr}(s)\Delta_{ar}(s{+}l{-}k)
                      +\Delta_{ra}(s)\Delta_{rr}(s{+}l{-}k)\Bigr]\, ,
 \nonumber\\
 \label{krrar}
  K&&_{rrar}(-k,k,l,-l)={\lambda^2\over 8}\int{d^4s\over (2\pi)^4}
 \\
   &&\quad\times \Bigl[\Delta_{ar}(s)\Delta_{rr}(s{+}l{-}k)
                 +\Delta_{rr}(s)\Delta_{ra}(s{+}l{-}k)\Bigr]\, ,
 \nonumber\\
  K&&_{rraa}(-k,k,l,-l) ={\lambda^2\over 8}\int{d^4s\over (2\pi)^4}
    \Bigl[\Delta_{rr}(s)\Delta_{rr}(s{+}l{-}k)
 \nonumber\\
     &&\quad +\Delta_{ra}(s)\Delta_{ar}(s{+}l{-}k)
             +\Delta_{ar}(s)\Delta_{ra}(s{+}l{-}k)\Bigr]\, .
 \label{krraa}
 \end{eqnarray}
 \end{mathletters}
Noticing that the $s^0$-integrals over terms of the type
$\Delta_{ra}(s)\Delta_{ra}(s{+}l{-}k)$ and 
$\Delta_{ar}(s)\Delta_{ar}(s{+}l{-}k)$ vanish because the poles of
these two terms are both on the same side of the real axis in the 
complex $s^0$ plane, we can make the following replacement: 
 \begin{eqnarray}
 \label{approx}
   &&\int{d^4s\over (2\pi)^4} \Bigl[\Delta_{ra}(s)\Delta_{ar}(s{+}l{-}k)
    +\Delta_{ar}(s)\Delta_{ra}(s{+}l{-}k)\Bigr]
 \nonumber\\
   &&={\int}{d^4s\over (2\pi)^4} \bigl[\Delta_{ra}(s){-}\Delta_{ar}(s)\bigr]
     \bigl[\Delta_{ar}(s{+}l{-}k){-}\Delta_{ra}(s{+}l{-}k)\bigr]
 \nonumber\\
   &&=\int{d^4s\over (2\pi)^4}\, \rho(s)\,\rho(s{+}l{-}k)\, .
 \end{eqnarray}
Inserting Eqs.~(\ref{kphi4}) into Eq.~(\ref{kernel}) and using 
Eqs.~(\ref{FDT1}), (\ref{density}) and (\ref{approx}), we express the 
kernel of the BS equation as
 \begin{equation}
 \label{kerphi4}
  {\cal K}^{(\phi^4)}(-k,k,l,-l)={\lambda^2\over 2}\int d[s_1,s_2]\,,
 \end{equation}
introducing for later convenience the shorthand notation
 \begin{eqnarray}
 \label{notation}
  &&\int d[s_1, s_2] \equiv -{{1+n(l^0)}\over {1+n(k^0)}}
    \int{d^4s_1\over (2\pi)^4}{d^4s_2\over (2\pi)^4}
    \bigl(1{+}n(s_1^0)\bigr)\rho(s_1)
 \nonumber\\
  &&\qquad \times \bigl(1{+}n({-}s_2^0)\bigr) \rho(-s_2)\,
    (2\pi)^4\delta^4(s_1{-}s_2{+}l{-}k)\, .
 \end{eqnarray}
Substituting this into Eq.(\ref{Dpion}) we obtain
 \begin{eqnarray}
 \label{Dpionphi4}
    &&D_{\pi}(k)=I_{\pi}(k) +{\lambda^2\over 4}
      {\int}{d^4l\over (2\pi)^4} {1{+}n(l^0)\over 1{+}n(k^0)}\,
      {\rho(l)\,D_{\pi}(l)\over{\rm Im\, }\Sigma(l)} 
 \nonumber\\
  &&\quad\times   
    {\int}{d^4s\over (2\pi)^4} 
    n(s^0)\rho(s) \bigl(1{+}n(s^0{+}k^0{-}l^0)\bigr)\rho(s{+}k{-}l) \, .
 \end{eqnarray}
The shear viscosity is obtained by inserting the solution of this
integral equation into Eq.~(\ref{eta5}). This result coincides with 
that obtained by Jeon \cite{Jeon2} in the ITF.

\subsection{Massive $g\phi^3+\lambda\phi^4$ theory}
\label{sec3b}
\subsubsection{The $\phi$ self energy}
\label{sec3b1}

\begin{figure}
 \epsfxsize 80mm \epsfbox{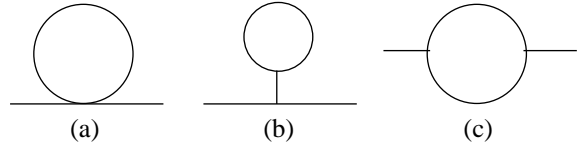}
 \vskip 0.4cm
 \caption{\label{F3}
  \small One-loop self-energy diagrams in $g\phi^3{+}\lambda\phi^4$ theory.}
 \end{figure}

\begin{figure}
 \epsfxsize 80mm \epsfbox{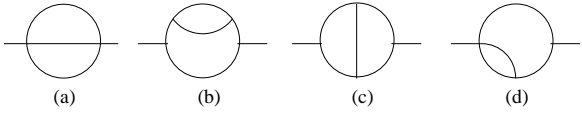}
 \vskip 0.4cm
 \caption{\label{F4}
  \small Two-loop self-energy diagrams in $g\phi^3{+}\lambda\phi^4$ theory.}
 \end{figure}

We now investigate the contributions arising from an additional cubic
interaction. For massless scalar fields, it renders the zero temperature
theory perturbatively unstable, causing a nonzero vacuum expectation 
value for $\phi$. To avoid this we introduce a nonzero mass $m$ for 
the scalar field and study \cite{Jeon2}
 \begin{equation}
 \label{lag1}
    {\cal L}_0 = {1\over 2} (\partial_{\mu}\phi)^2 - {m^2\over 2} \phi^2
    - {g\over 3!}\phi^3 - {\lambda \over 4!} \phi^4 + {\cal L}_c
 \end{equation}
in the weak-coupling limit $(g/m)^2{\,\lesssim\,}\lambda{\,\ll\,}1$.
$m$, $g$, and $\lambda$ denote the physical mass and coupling constants 
at zero temperature, and ${\cal L}_c$ denotes the counter terms required 
for renormalization. 
The one- and two-loop diagrams for the self-energy in this theory are 
shown in Figs.\,\ref{F3} and \ref{F4}.  
 
Diagrams \ref{F3}(a) and \ref{F3}(b) are independent of the external 
momentum which does not enter the loop; they generate a purely real 
contribution ${\cal O}(\lambda T^2{-}g^2T^2/m^2)$ to the scalar self 
energy. For $T{\,\ll\,}m/\sqrt{\lambda}$ this is much smaller than 
$m^2$, and the viscosity can thus be calculated using the 
zero-temperature mass $m$ in all propagators \cite{Jeon2}. For 
$T{\,\gtrsim\,}m/\sqrt{\lambda}$, the thermal self energy becomes 
comparable to $m^2$ and must be resummed, yielding quasi-particles 
with a thermal mass of order $m_{\rm th}^2{\,\sim\,}\lambda T^2$ 
\cite{Jeon2,Parwani,Wang4}. For $T{\,\gg\,}m/\sqrt{\lambda}$ the 
self-energy from diagram \ref{F3}(b) is then 
${\sim\,}{-}g^2T^2/m_{\rm th}^2$; due to the weak-coupling condition 
$g^2{\,\lesssim\,}\lambda m^2$ this is smaller than diagram \ref{F3}(a) 
by a factor $m^2/m_{\rm th}^2{\,\sim\,}m^2/(\lambda T^2){\,\ll\,}1$. 
At such high temperatures, the effects from the 4-point interaction 
thus dominate those from the 3-point interaction, and the theory becomes
effectively a massless $\lambda\phi^4$ theory as discussed in the previous
subsection. We thus concentrate on the interesting temperature range 
$T{\,\sim\,}m/\sqrt{\lambda}$ where on the one hand mass resummation 
is required but on the other hand $\phi^3$ interaction effects are not 
yet negligible.

The lowest order contribution to the imaginary part of the scalar self 
energy arises from diagram \ref{F3}(c). It vanishes on-shell (an on-shell 
particle cannot decay into two on-shell particles with the same thermal 
mass) but is non-zero and ${\cal O}(g^2)$ for off-shell momenta. Its real 
part is smaller than that of Figs.~\ref{F3}(a,b) and does not contribute 
to the leading-order thermal mass \cite{Jeon2}. The imaginary contribution 
vanishes for the side rails of the multi-ladder diagrams in the 
``nearly pinching poles'' approximation which forces the side rail 
momenta approximately on-shell, but it plays an important role for
single-line rungs in the BS equation generated by the exchange of a
single scalar field (see Fig.~\ref{F5}(b) below).

The leading order contribution to the {\em on-shell} imaginary
part of the self-energy arises from the two-loop diagrams in
Fig.\,\ref{F4}. As shown in detail in \cite{Jeon2}, it is 
${\cal O}(\lambda^2 T^2)$ (this includes the contributions from 
diagrams \ref{F4}(b-c) in the weak-coupling limit 
$(g/m)^2{\,\sim\,}\lambda{\,\ll\,}1$). It results in an 
${\cal O}(1/(\lambda^2 T))$ 
thermal lifetime for the quasi-particles \cite{Jeon2,Parwani,Wang4}. Just 
as in the simpler $\phi^4$ theory, the frequency integral over pairs of 
propagators from the side rails of the multi-ladder diagram sharing the 
same loop momentum is proportional to the quasi-particle lifetime. The 
corresponding factor $1/\lambda^2{\,\sim\,}m^2/(g^2\lambda){\,\sim\,}(m/g)^4$ 
compensates for the extra coup\-ling constants from the vertices associated 
with the rung. Thus again an infinite number of ladder diagrams contributes 
to the leading order shear viscosity; they will be resummed through the 
BS equation.

\subsubsection{The BS kernel for $g\phi^3+\lambda\phi^4$ theory}
\label{sec3b2}

\begin{figure}
 \epsfxsize 80mm \epsfbox{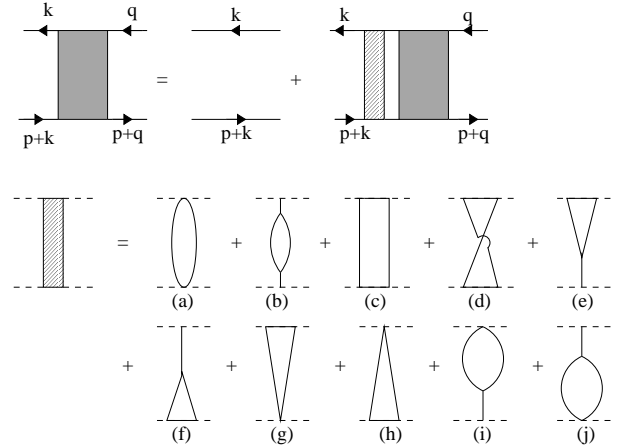}
 \vskip 0.4cm
 \caption{\label{F5}
  \small Graphic representation of the Bethe-Salpeter equation 
  for the 4-point Green function in $g\phi^3{+}\lambda\phi^4$ theory.
  Compared to the pure $\lambda\phi^4$ theory shown in Fig.~\ref{F2}, 
  the kernel giving the leading order contribution to the shear viscosity
  (shown in the second and third line) is more elaborate. With 
  $(g/m)^2{\,\sim\,}\lambda$, all contributions to the kernel are of order 
  $\lambda^2$.}
 \end{figure}

The various contributions to the kernel of the BS equation for the 
leading order shear viscosity in $g\phi^3 +\lambda\phi^4$ theory are
illustrated in Fig.\,\ref{F5}. Whereas for the side rails of the ladder
{\em full} (i.e. at least two-loop resummed) propagators must be used,
in order to regulate the pinch singularity in the integral over the
side rail momentum, it is sufficient at leading order to use bare
propagators in the contributions to the rungs of the ladder shown in 
Fig.~\ref{F5} \cite{Jeon2}. We will nonetheless formally evaluate them 
with full propagators, in keeping with the spirit of our skeleton 
diagram expansion, even though we will not make use of the implied higher
order corrections which, for a consistent treatment, would also require
the resummation of non-ladder diagrams and ladders with crossed rungs.

Diagram \ref{F5}(a) corresponds to the pure $\phi^4$ theory discussed 
in the previous subsection. For pure $\phi^3$ theory, one might think 
that the simplest kernel for the ladder diagrams should arise from the 
single straight line rung (one-particle exchange) (not shown in 
Fig.~\ref{F5}). Using naive power counting, this straight-line kernel 
would contribute only two powers of $g$ from the explicit vertices which 
cannot even compensate for the ``nearly pinching poles'' contribution 
${\sim\,}1/\lambda^2{\,\sim\,}(m/g)^4$ from the frequency integral 
over the pair of side rail propagators with the same loop momentum. If 
this were correct, in the weak-coupling limit each further single-line 
rung on the ladder would make the shear viscosity more singular. 
Fortunately, this fear is unnecessary: due to the nearly pinching 
poles from the pairs of rail propagators, the loop integrals are 
dominated by almost on-shell rail momenta. For non-zero values of 
the overall loop momentum $k$ in Eq.~(\ref{eta3}) this forces all 
rung momenta off-shell. In the $(r,a)$ basis, only vertices with an 
odd number of $a$ indices are non-zero; together with 
$\Delta_{aa}(k){\,=\,}0$ this implies that the kernel from the straight 
line rung is proportional to the spectral density of the $\phi$
propagator (see Eq.~(\ref{kfig5b}) below). For off-shell momenta, 
this spectral density vanishes at tree level, so the single straight 
line rung with a bare exchanged propagator does not contribute.

A non-zero contribution to the kernel from single particle exchange
thus requires at least 1-loop accuracy for the propagator of the 
single-line rung; this is shown in Fig.~\ref{F5}(b). As discussed in 
the preceding subsection, the loop in diagram \ref{F3}(c) contributes 
an imaginary part of order $g^2$ to the single-particle self-energy and 
spectral density which is non-zero for off-shell momenta. Together with 
the two factors $g$ from the vertices on the side rail this renders the 
kernel from Fig.~\ref{F5}(b) ${\cal O}\bigl((g/m)^4\bigr)$ and thus of 
the same order as all other diagrams in Fig.~\ref{F5} (which involve the 
exchange of more than one particle). Note that, whereas the exchange of 
a single bare propagator contributes zero to the BS kernel, its once 
iterated version, shown in Fig.~\ref{F5}(c), is non-zero and again of 
order ${\cal O}\bigl((g/m)^4\bigr)$ \cite{Jeon2}.

\subsubsection{The individual contributions to the kernel}
\label{sec3b3}

The kernel from diagram \ref{F5}(a) is given by Eq.~(\ref{kerphi4}):
 \begin{equation}
 \label{kfig5a+}
   {\cal K}^{(a)}(-k,k,l,-l)={\lambda^2\over 2}\int d[s_1,s_2].
 \end{equation}
The other contributions are calculated in a similar way. We show the
calculational process in some detail for dia\-grams \ref{F5}(b) and 
\ref{F5}(c), but only mention a few key steps and list the final 
results for diagrams \ref{F5}(d-j). 

In the $(r,a)$ basis the bare 3-point vertex is given by 
 \begin{equation}
 \label{gcoupling}
  g_{\alpha_1\alpha_2\alpha_3}=
  {g\over {2^{3/2}}}\Bigl(1-(-1)^{n_a}\Bigr)\, ,
 \end{equation}
where $n_a$ is the number of $a$ indices among 
$(\alpha_1,\alpha_2,\alpha_3)$. Explicitly this reads 
$g_{rrr}{\,=\,}g_{raa}{\,=\,}g_{ara}{\,=\,}g_{aar}{\,=\,}0$,
$g_{aaa}=g_{rra}=g_{rar}=g_{arr}=g/\sqrt{2}$.

For the single-particle exchange Fig.~\ref{F5}(b) the general expression 
for the kernel can be written as
 \begin{eqnarray}
 \label{kerfig5b}
   &&K^{(b)}_{\alpha_1\beta_1\alpha_3\beta_3}(-k,k,l,-l)
 \nonumber\\
   &&\quad=-i g_{\alpha_1\alpha_2\alpha_3}
    g_{\beta_1\beta_2\beta_3}
    \Delta_{\beta_2\alpha_2}(k{-}l)
 \end{eqnarray}
where $\Delta$ is the full single-particle propagator. With 
Eq.~(\ref{gcoupling}) we get
 \begin{mathletters}
 \label{f5}
 \begin{eqnarray}
   K^{(b)}_{rrra}(-k,k,l,-l)&=& -i {g^2\over 2} \Delta_{ra}(k{-}l)\, ,
 \label{krrrafig5b}\\
   K^{(b)}_{rrar}(-k,k,l,-l)&=& -i {g^2\over 2} \Delta_{ar}(k{-}l)\, ,
 \label{krrarfig5b}\\
   K^{(b)}_{rraa}(-k,k,l,-l)&=& -i {g^2\over 2} \Delta_{rr}(k{-}l)\, .
 \label{krraafig5b}
 \end{eqnarray}
 \end{mathletters}
Inserting these into Eq.~(\ref{kernel}) and using Eqs.~(\ref{FDT1}),
(\ref{density}) and (\ref{pinch}) we obtain
 \begin{eqnarray}
 \label{kfig5b}
   &&{\cal K}^{(b)}(-k,k,l,-l)
     =-{g^2\over 2}[N(l^0)+N(k^0{-}l^0)]\,\rho(k{-}l)
 \nonumber\\
    &&\ \  =-g^2[N(l^0){+}N(k^0{-}l^0)]
    \bigl|\Delta_{ra}(k{-}l)\bigr|^2 \,{\rm Im}\,\Sigma(k{-}l)\, .
 \end{eqnarray}
As promised, the kernel for diagram \ref{F5}(b) is proportional to the 
spectral density (or the imaginary part of the self-energy). For off-shell 
$k{-}l$ the lowest order contribution to Im\,$\Sigma$ is given by 
diagram \ref{F3}(c):
 \begin{eqnarray}
 \label{imS}
   &&{\rm Im}\,\Sigma(k{-}l)={1\over 2i}
   \Bigl(\Sigma_{ra}(k{-}l)-\Sigma_{ar}(k{-}l)\Bigr)
 \nonumber\\
    &&\quad ={g^2\over 8}{\int}{d^4s\over (2\pi)^4}
    \Bigl\{\Delta_{rr}(s)
           \bigl[\Delta_{ar}(s{+}k{-}l){-}\Delta_{ra}(s{+}k{-}l)\bigr]
 \nonumber\\
    &&\qquad\qquad +\bigr[\Delta_{ra}(s){-}\Delta_{ar}(s)\bigr]
             \Delta_{rr}(s{+}k{-}l)\Bigr\}\, .
 \end{eqnarray}
Using Eqs.~(\ref{FDT1}), (\ref{density}) and the identity
 \begin{eqnarray}
 \label{disid}
 &&n(z_1 {+} z_2 {+}\cdots {+} z_n)=
 \\
 &&{{n(z_1)n(z_2)\cdots n(z_n)}\over
   {[1{+}n(z_1)][1{+}n(z_2)]\cdots [1{+}n(z_n)]
    {-}n(z_1)n(z_2)\cdots n(z_n)}}\,,
 \nonumber
 \end{eqnarray}
we can insert Eq.~(\ref{imS}) back into Eq.~(\ref{kfig5b}). With a little
algebra we find that the result can be written in the form
 \begin{eqnarray}
 \label{kfig5b+}
  {\cal K}^{(b)}(-k,k,l,-l) ={g^4\over 2}\int d[s_1, s_2]\,
    \bigl|\Delta_{ra}(s_1{-}s_2)\bigr|^2\,,
 \end{eqnarray}
where we used the $\delta$-function in Eq.~(\ref{notation}) to replace
$k{-}l{\,=\,}s_1{-}s_2$.

For diagram \ref{F5}(c) the kernel can be expressed as
 \begin{eqnarray}
 \label{kerfig5c}
   &&K^{(c)}_{\alpha_1\beta_1\gamma_3\delta_3}(-k,k,l,-l)
   =g_{\alpha_1\alpha_2\alpha_3}
    g_{\beta_1\beta_2\beta_3}
    g_{\gamma_1\gamma_2\gamma_3}
    g_{\delta_1\delta_2\delta_3}
 \nonumber\\
   &&\quad\times
    \int{d^4s\over (2\pi)^4}
    \Delta_{\beta_2\alpha_2}(s)\Delta_{\gamma_2\delta_2}(s{+}l{-}k)
 \nonumber\\
   &&\qquad\qquad\quad\!\times\,
    \Delta_{\beta_3\delta_1}(k{-}s)\Delta_{\gamma_1\alpha_3}(k{-}s)
    \,.
 \end{eqnarray}
With Eqs.~(\ref{gcoupling}), (\ref{daa}), and (\ref{FDT1}) and dropping 
the terms involving $\Delta_{ra}(k{-}s)\Delta_{ra}(k{-}s)$ and
$\Delta_{ar}(k{-}s)\Delta_{ar}(k{-}s)$, we obtain
 \begin{mathletters}
 \label{fig5c}
 \begin{eqnarray}
   &&K^{(c)}_{rrra}(-k,k,l,-l)
 \nonumber\\
   &&\quad ={g^4\over 4}{\int}{d^4s\over (2\pi)^4}
    \bigl|\Delta_{ra}(k{-}s)\bigr|^2\Delta_{ar}(s{+}l{-}k)
 \nonumber\\
   &&\qquad\ \times \Bigl[ \Delta_{rr}(s) + N(k^0{-}s^0)
      \Bigl(\Delta_{ra}(s){-}\Delta_{ar}(s)\Bigr)\Bigr]\, ,
 \label{krrrafig5c}\\
   &&K^{(c)}_{rrar}(-k,k,l,-l)
 \nonumber\\
   &&\quad ={g^4\over 4}{\int}{d^4s\over (2\pi)^4}
    \bigl|\Delta_{ra}(k{-}s)\bigr|^2\Delta_{ra}(s{+}l{-}k)
 \nonumber\\
   &&\qquad\ \times \Bigl[\Delta_{rr}(s) + N(k^0{-}s^0)
      \Bigl(\Delta_{ra}(s){-}\Delta_{ar}(s)\Bigr)\Bigr]\, ,
 \label{krrarfig5c}\\
   &&K^{(c)}_{rraa}(-k,k,l,-l)
  \nonumber\\
   &&\quad ={g^4\over 4}{\int}{d^4s\over (2\pi)^4}
    \bigl|\Delta_{ra}(k{-}s)\bigr|^2 \Delta_{rr}(s{+}l{-}k)
 \nonumber\\
   &&\qquad\ \times \Bigl[\Delta_{rr}(s) + N(k^0{-}s^0)
      \Bigl(\Delta_{ra}(s){-}\Delta_{ar}(s)\Bigr)\Bigr]\, .
 \label{krraafig5c}
 \end{eqnarray}
 \end{mathletters}
Inserting these expressions into Eq.~(\ref{kernel}) and using 
Eqs.~(\ref{FDT1}), (\ref{density}) and (\ref{disid}) we find
 \begin{eqnarray}
 \label{kfig5c+}
  &&{\cal K}^{(c)}(-k,k,l,-l) =g^4\int d[s_1, s_2]\,
    \bigl|\Delta_{ra}(k{+}s_2)\bigr|^2
 \nonumber\\
  &&\quad ={g^4\over 2}\int d[s_1, s_2]\,
    \Bigl[\bigl|\Delta_{ra}(s_1{-}k)\bigr|^2
         +\bigl|\Delta_{ra}(s_1{+}l)\bigr|^2 \Bigr]\,.
 \end{eqnarray}
In the second equality we used the $\delta$-function in Eq.(\ref{notation})
to change variables $s_1{\,\leftrightarrow\,}{-}s_2$ and replace
$s_2{+}k{\,\to\,}s_1{+}l$.

Repeating this procedure we derive also the remaining contributions 
to the BS kernel: 
 \begin{mathletters}
 \label{rest}
 \begin{eqnarray}
  &&{\cal K}^{(d)}(-k,k,l,-l)
 \nonumber\\
   &&\quad =g^4{\int}d[s_1, s_2]\,
    {\rm Re}\Bigl[\Delta_{ra}(s_1{+}l)\Delta_{ar}(s_1{-}k)\Bigr]\, ,
 \label{kfig5d+}\\
  &&{\cal K}^{(e)}(-k,k,l,-l)
 \nonumber\\
   &&\quad =g^4{\int} d[s_1, s_2]\,
    {\rm Re}\Bigl[\Delta_{ra}(s_1{-}k)\Delta_{ar}(s_1{-}s_2)\Bigr]\, ,
 \label{kfig5e+}\\
  &&{\cal K}^{(f)}(-k,k,l,-l)
 \nonumber\\
   &&\quad =g^4{\int} d[s_1, s_2]\,
    {\rm Re}\Bigl[\Delta_{ra}(s_1{+}l)\Delta_{ar}(s_1{-}s_2)\Bigr]\, ,
 \label{kfig5f+}\\
  &&{\cal K}^{(g)}(-k,k,l,-l)
     =\lambda g^2{\int}d[s_1, s_2]\,{\rm Re\,}\Delta_{ra}(s_1{-}k)\, ,
 \label{kfig5g+}\\
  &&{\cal K}^{(h)}(-k,k,l,-l)
    =\lambda g^2{\int} d[s_1, s_2]\,{\rm Re\,}\Delta_{ra}(s_1{+}l)\, ,
 \label{kfig5h+}\\
  &&{\cal K}^{(i)}(-k,k,l,-l)={\cal K}^{(j)}(-k,k,l,-l)
 \nonumber\\
   &&\quad ={{\lambda g^2}\over 2}{\int}d[s_1, s_2]\,
    {\rm Re\,}\Delta_{ra}(s_1{-}s_2)\, .
 \label{kfig5i+}
 \end{eqnarray}
 \end{mathletters}
In key to obtaining these results is, beyond using the relations already 
mentioned above, a judicious redefinition of integration variables in
order to bring them all into the same form. We also exploited the equality
$\Delta_{\alpha_1\alpha_2}(-k){\,=\,}\Delta_{\alpha_2\alpha_1}(k)$. 
Noting that diagrams \ref{F5}(e), \ref{F5}(g) and \ref{F5}(i) can be 
obtained by turning diagrams \ref{F5}(f), \ref{F5}(h) and \ref{F5}(j)
upside down, we can use the following relations to simplify the
calculation of the last three: The symmetry of the diagrams (e,f) 
gives
 \begin{mathletters}
 \label{sym}
 \begin{eqnarray}
    K^{(f)}_{rrra}(-k,k,l,-l) &=& K^{(e)}_{rrar}(k,-k,-l,l)\, ,
 \label{sym1}\\
    K^{(f)}_{rrar}(-k,k,l,-l) &=& K^{(e)}_{rrra}(k,-k,-l,l)\, ,
 \label{sym2}\\
    K^{(f)}_{rraa}(-k,k,l,-l) &=& K^{(e)}_{rraa}(k,-k,-l,l)\, .
 \label{sym3}
 \end{eqnarray}
 \end{mathletters}
Substituting this into Eq.~(\ref{kernel}) and using 
$N(-l^0)={-}N(l^0)$ we obtain
 \begin{equation}
 \label{symfig5f}
 {\cal K}^{(f)}(-k,k,l,-l)={\cal K}^{(e)}(k,-k,-l,l)\, .
 \end{equation}
${\cal K}^{(f)}(-k,k,l,-l)$ is thus obtained from 
${\cal K}^{(e)}(-k,k,l,-l)$ by simply replacing $k{\,\to\,}{-}k$ and 
$l{\,\to\,}{-}l$. Similarly we find
 \begin{mathletters}
 \begin{eqnarray}
 {\cal K}^{(h)}(-k,k,l,-l)&=&{\cal K}^{(g)}(k,-k,-l,l)\, ,
 \label{symfig5h}\\
 {\cal K}^{(j)}(-k,k,l,-l)&=&{\cal K}^{(i)}(k,-k,-l,l)\, .
 \label{symfig5j}
 \end{eqnarray}
 \end{mathletters}
The last equation implies the equality of the kernels from diagrams
\ref{F5}(i) and \ref{F5}(j).

\subsubsection{The final result}
\label{sec3b4}

We can now add the contributions (\ref{kfig5a+}), (\ref{kfig5b+}),
(\ref{kfig5c+}), and (\ref{rest}) to the BS kernel for the shear 
viscosity in $g\phi^3+\lambda\phi^4$ theory:
 \begin{eqnarray}
 \label{kerphi4+3}
  &&{\cal K}^{(g\phi^3{+}\lambda\phi^4)}(-k,k,l,-l)
    ={1\over 2}{\int}d[s_1, s_2]\,
 \\
  &&\times \Bigl|\lambda +g^2 \Bigl(\Delta_{ra}(s_1{+}l)
    +\Delta_{ra}(s_1{-}k) +\Delta_{ra}(s_1{-}s_2)\Bigr)\Bigr|^2\, .
 \nonumber
 \end{eqnarray}
After accounting for the different metric convention for the ITF propagator 
used in Ref.~\cite{Jeon2} which results in a relative minus sign to the 
propagator defined here in Eq.~(\ref{fullret}), this result agrees
completely with the one obtained by Jeon in Eq.~(4.43) of Ref.~\cite{Jeon2}.

The integrand in Eq.~(\ref{kerphi4+3}) corresponds to the square
of the tree-level two-body "scattering amplitude" in finite
temperature $g\phi^3+\lambda\phi^4$ theory. Starting from the 
result (\ref{kerphi4}) for pure $\phi^4$ theory, one can obtain
the BS kernel for $g\phi^3+\lambda\phi^4$ theory by simply replacing 
the tree-level two-body scattering amplitude $\lambda$ in $\phi^4$ 
theory by the corresponding more elaborate amplitude in 
$g\phi^3{+}\lambda\phi^4$ theory. Substituting Eq.~(\ref{kerphi4+3}) 
into the BS integral equation (\ref{Dpion}) and solving the latter 
(which for pure $\phi^4$ theory was done in \cite{Jeon2} and will not
be repeated here), one arrives at the nonperturbative result for 
leading order shear viscosity. It is of order 
$\eta{\,\sim\,}T^3/\lambda^2{\,\sim\,}T^3(m/g)^4$ 
\cite{Jeon1,Jeon2,Wang2,Wang4}.

\section{Summary and conclusions}
\label{sec4}
\vspace*{-1.32cm}
For the shear viscosity in hot plasmas a nonperturbative calculation 
is needed, resumming an infinite number of ladder diagrams which all
contribute at leading order. Using a real-time approach based on the 
CTP formalism \cite{Schwinger,Bakshi,Keldysh}, we formulated this task 
in terms of a BS integral equation for the 4-point Green function. In 
this approach, the BS equation is a tensor equation which couples the 
thermal components of the 4-point function among each other. We showed 
that, at leading order, it can be decoupled by using the $(r,a)$ basis 
of Chou {\em et al.} \cite{Chou} rather than the familiar 
Schwinger-Keldysh basis \cite{Schwinger,Bakshi,Keldysh}. In 
the $(r,a)$ basis, the leading order contribution to the shear 
viscosity $\eta$ is given by the single $G_{aarr}$ component of the 
4-point function. Decoupling of its BS equation makes abundant use 
of the Generalized Fluctuation-Dissipation Theorem for non-linear 
response functions \cite{Wang3}. 

The present derivation of the leading order shear viscosity is 
considerably more compact than the previous ITF calculation by 
Jeon \cite{Jeon2}. It differs from the calculation of Carrington 
{\it et al.} \cite{CHK} by following a strict organization in terms
of skeleton diagrams with full propagators and bare vertices which we
find conceptually more appealing. This approach clarifies a disagreement
between Refs.~\cite{Jeon2} and Ref.~\cite{CHK} as to the type of diagrams
to be resummed in a complete leading order calculation: we do not find a 
need for including any diagrams beyond those already considered by Jeon 
\cite{Jeon2}. Within our approach we first derived a {\em general} 
expression for the shear viscosity which describes the high temperature 
limit of {\em all} scalar field theories, and then evaluated the specific 
forms of the BS integral kernel for the $\lambda\phi^4$ and 
$g\phi^3{+}\lambda\phi^4$ interaction Lagrangians. Our results fully 
reproduce those obtained by Jeon \cite{Jeon2} in the ITF, providing an 
important check and simplification of that hallmark calculation. For 
$(g/m)^2{\,\sim\,}\lambda{\,\ll\,}1$, both theories give a leading order 
result for the shear viscosity of order $\eta{\,\sim\,}T^3/\lambda^2$.  

Due to the need for resummation, the diagrammatic approach to transport
coefficients in hot field theories, based on Kubo formulae, has 
developed a reputation for being extremely demanding and cumbersome.
This is particularly true for the ITF formalism with its need for
analytical continuation and the corresponding complicated cutting rules 
\cite{Jeon2}. As a result, an alternative approach based on the 
Boltzmann equation in kinetic theory has recently become more popular 
\cite{Jeon:1995zm}. At this point, the only published calculation of 
the leading-order shear viscosity in hot gauge theories \cite{AMY} 
is based on that approach. But these calculations are not trivial 
either, and they cannot fully replace the kind of intuitive insight into 
the underlying physical mechanisms which is provided by a diagrammatic 
analysis. A recent computation \cite{MartinezResco:2000pz} of the color 
conductivity in QCD, using the ITF diagrammatic approach, sucessfully 
reproduced an earlier result \cite{Arnold:1998cy} based on an effective 
Boltzmann equation. A still unpublished calculation \cite{resco2001} of 
the shear viscosity in hot QED and QCD, using the same diagrammatic ITF 
approach, differs slightly from the result in \cite{AMY}, while a
recent paper by Valle Basagoiti \cite{VB02} which claims agreement
between the Kubo formula and Boltzmann equation approaches employs
a resummed 3-point vertex which apparently does not satisfy the Ward
identity \cite{resco2002}. (The violation of the Ward identity may not
affect the leading logarithmic result, but this requires further study.)   
It is our hope that the more compact real-time formalism in the $(r,a)$ 
basis developed here may help to resolve the existing discrepancies and 
lead to a revival of resummed perturbation theory as a tool for computing 
relativistic transport coefficients. We also believe that it may serve 
as a suitable starting point for perturbative calculations of 
non-equilibrium transport phenomena, although away from thermal equilibrium 
the simplifications provided by the Fluctuation-Dissipation Theorem no 
longer work and hence additional complications must be expected. 

\acknowledgments

We thank Gert Aarts for valuable comments on the manuscript.
This work was supported by the National Natural Science Foundation
of China (NSFC) under project Nos. 19928511, 19945001 and
10135030, and by the U.S. Department of Energy under contract 
DE-FG02-01ER41190. E.W. thanks the Nuclear Theory Group in the 
Department of Physics at the Ohio State University for their 
hospitality during the completion of this work.


\end{multicols}

\begin{references}

\bibitem{Zubarev}
   D. N. Zubarev, {\it Nonequilibrium Statistical Thermodynamics} 
   (Plenum, New York, 1974).
\bibitem{Hosoya}
  A. Hosoya, M.-A. Sakagami, and M. Takao, Ann. Phys. (N.Y.) {\bf 154}, 229
  (1984).
\bibitem{Jeon1}
  S. Jeon, Phys. Rev. D {\bf 47}, 4586 (1993).
\bibitem{Wang1}
  E. Wang, U. Heinz, and X. Zhang, Phys. Rev. D {\bf 53}, 5978 (1996).
\bibitem{Jeon2}
  S. Jeon, Phys. Rev. D {\bf 52}, 3591 (1995).
\bibitem{Wang2}
  E. Wang and U. Heinz, Phys. Lett. B {\bf 471}, 208 (1999).
\bibitem{CHK}
  M.E. Carrington, Hou Defu, and R. Kobes, Phys. Rev. D {\bf 62}, 
  025010 (2000).
\bibitem{Schwinger}
  J. Schwinger, J. Math. Phys. {\bf 2}, 407 (1961).
\bibitem{Bakshi}
   K.T. Mahanthappa, Phys. Rev. {\bf 126}, 329 (1962); 
   P.M. Bakshi and K.T. Mahanthappa, J. Math. Phys. {\bf 4}, 1 and 12 (1963).
\bibitem{Keldysh}
  L.V. Keldysh, Sov. Phys. JETP {\bf 20}, 1018 (1965).
\bibitem{Chou}
  K.-C. Chou, Z.-B. Su, B.-L. Hao, and L. Yu, Phys. Rep. {\bf 118}, 1 (1985).
\bibitem{Wang3}
  E. Wang and U. Heinz, {\it A Generalized  Fluctuation-Dissipation 
  Theorem for Nonlinear Response Functions}, Phys. Rev. D, in press 
  [hep-th/9809016].
\bibitem{CHS}
  M.E. Carrington, Hou Defu, and J.C. Sowiak, Phys. Rev. D {\bf 62}, 
  065003 (2000).
\bibitem{AMY}
  P.~Arnold, G.~D.~Moore, and L.~G.~Yaffe, JHEP {\bf 0011} (2000) 001.
\bibitem{Callen}
  H.B. Callen and T.A. Welton, Phys. Rev. {\bf 83}, 34 (1951).
\bibitem{fn1}
  This pinch singularity is physical and reflects the absence of 
  collisions (infinite time between scatterings) in the noninteracting 
  theory. In kinetic theory within the relaxation time approximation, 
  infinite scattering time means infinite relaxation time and therefore 
  infinite viscosity \cite{Zubarev}. To obtain a finite viscosity, 
  scattering processes must be taken into account. These give the plasma
  particles a non-zero collisional width which shifts the poles of the 
  single particle propagator away from the real axis, thereby regulating 
  the pinch singularity and rendering the limits of zero momentum and 
  energy in Eqs. (\ref{eta1}) and (\ref{eta2}) well-defined and independent 
  of the order in which they are taken. Following Jeon \cite{Jeon1,Jeon2}, 
  we account for this need for a finite collisional width by organizing 
  the calculation in terms of skeleton diagrams involving {\em full} 
  propagators. Extracting the leading order contribution to the viscosity
  then amounts to specifying for each line in the skeleton diagram 
  individually the required accuracy for the single particle self-energy 
  in powers of the coupling constant.
\bibitem{Parwani}
  R.R. Parwani, Phys. Rev. D {\bf 45}, 4695 (1992).
\bibitem{Wang4}
  E. Wang and U. Heinz, Phys. Rev. D {\bf 53}, 899 (1996).
\bibitem{Guerin}
  F. Guerin, eprint archive hep-ph/0105313 and hep-ph/0111020.
\bibitem{Kubo}
   R. Kubo, J. Phys. Soc. Japan {\bf 12}, 570 (1957);  P.C. Martin and
   J. Schwinger, Phys. Rev. {\bf 115}, 1432 (1959).
\bibitem{Lehmann}
  H. Lehmann, K. Symanzik, and W. Zimmermann, Nuovo Cimento {\bf 6}, 319
 (1957).
\bibitem{Jeon:1995zm}
  S.~Jeon and L.G.~Yaffe, Phys. Rev. D {\bf 53}, 5799 (1996);
  P.~Arnold and L.G.~Yaffe, {\em ibid.} {\bf 62}, 125014 (2000).
\bibitem{MartinezResco:2000pz}
  J.M.~Martinez Resco and M.A.~Valle Basagoiti,
  Phys. Rev. D {\bf 63} (2001) 056008.
\bibitem{Arnold:1998cy}
  P.~Arnold, D.T.~Son, and L.G.~Yaffe, Phys. Rev. D {\bf 59}, 105020 (1999).
\bibitem{resco2001} 
  J.M.~Martinez Resco, PhD thesis, unpublished.
\bibitem{VB02}
  M.A.~Valle Basagoiti, hep-ph/0204334.
\bibitem{resco2002} 
  J.M.~Martinez Resco, private communication.

\end{references}
\end{document}